\documentclass[journal]{IEEEtran}
\usepackage{graphicx}
\graphicspath{{figures/}}
\usepackage{cite}
\usepackage{stmaryrd}
\usepackage{pifont}
\usepackage{amsfonts}
\usepackage{mathrsfs}
\usepackage{amssymb}
\usepackage{amsmath}
\usepackage{dsfont}
\usepackage{subfigure}
\usepackage{xcolor}
\usepackage{epstopdf}
\newcounter{MYtempeqncnt}
\begin{document}
\title{An Energy-Efficient VCO-Based Matrix Multiplier Block to Support On-Chip Image Analysis}

\author{Imon Banerjee  and Arindam Sanyal
\thanks{I. Banerjee is with the Laboratory of Quantitative Imaging, Stanford University School of Medicine, Stanford 94305, CA, USA. 
(e-mail: imonb@stanford.edu)}
\thanks{A. Sanyal is with Department of Electrical Engineering, The State University of New York at Buffalo, Buffalo 14260, NY,USA. (email: arindams@buffalo.edu)}
}
\maketitle

\begin{abstract}
	Images typically are represented as uniformly sampled data in the form of matrix of pixels/voxels. Therefore, matrix multiply-and-accumulate (MAC) forms the core of most state-of-the-art image analysis algorithms. While digital implementation of MAC has generally been the preferred approach, high power consumption is an impediment to adopting it for medical image analysis. In this work, we present a time-domain signal processing architecture which performs MAC operations with 7bit accuracy while consuming 400X lower energy than digital implementation. The proposed  architecture performs analog computation using mostly digital circuits and is suitable for scaled CMOS technologies. The proposed time-domain MAC architecture is expected to play a central role in empowering the advancement of various on-chip image analysis operations. 
\end{abstract}


\section{Introduction}
In the digital data era, 2D/3D image analysis operations (e.g. registration, feature calculation, interpolation, fusion) are the core processing block of a wide range of automated systems, including computer aided diagnosis (CAD) \cite{1,2}. For example, in the modern CAD systems, an important image analysis operation is the co-registration of positron emission tomography (PET) image and magnetic resonance image (MRI) which combines functional information from PET images with anatomical information in MR images. The efficient co-registration of PET and MRI images can pave the way for a better understanding of physiological and disease mechanisms in pre-clinical and clinical settings.

\begin{figure*}[tbh!]
	\centering{\includegraphics[width=0.9\textwidth]{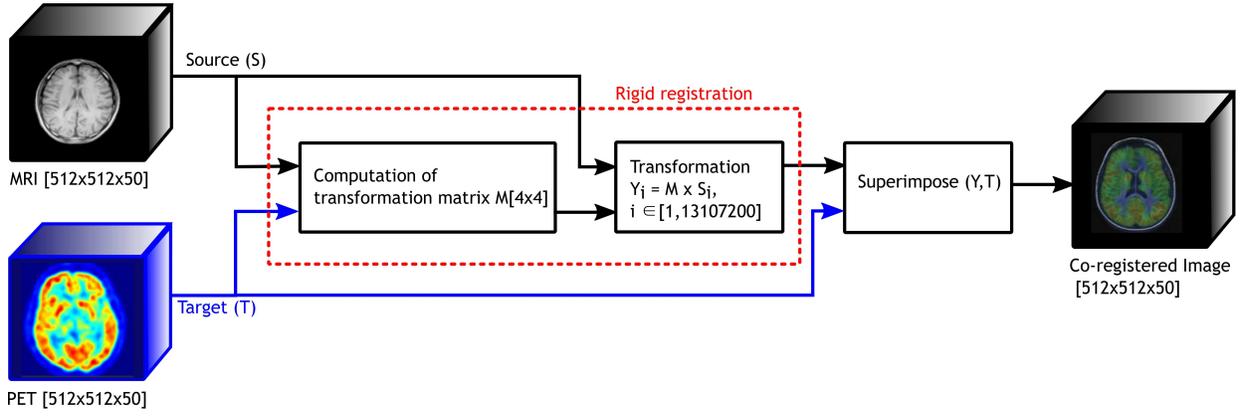}}
	\caption{A workflow diagram of MRI and PET brain image co-registration. }
	\label{fig0}
	\vspace{0mm}
\end{figure*}

The co-registration algorithm applies rigid registration~\cite{Goshtasby} to map each voxel $(x_1,y_1,z_1)$ in the MR image into the co-ordinate space of PET image $(x_2, y_2,z_2)$, provided that the images of both modalities are acquired from the same subject and the scanning processes have not introduced nonrigid spatial transformations. To illustrate the algorithmic facets, we present a workflow diagram in Fig. \ref{fig0} that takes MRI and PET images as inputs and uses rigid registration to compute the co-registered MRI/PET image. 
The rigid registration is done through multiply-and-accumulate (MAC) operations performed on the original voxel locations in 3D space (x,y,z) with $4\times4$ translation matrix ($U$) which represents translations($t$) and scaling($S$) and $4\times4$ rotation matrices ($R$) which represent rotation (see (\ref{eqn_res})). 

\begin{figure*}[!t]
\small
\setcounter{MYtempeqncnt}{\value{equation}}
\setcounter{equation}{0}
\begin{equation}
\label{eqn_res}
U = \left| \begin{array}{cccc} S_x & 0 & 0 & 0 \\0 & S_y & 0 & 0\\0 & 0 & S_z & 0\\t_x & t_y & t_z & 1 \end{array} \right|;
R_x = \left| \begin{array}{cccc} 1 & 0 & 0 & 0 \\0 & \cos \theta & -\sin \theta & 0\\0 & \sin \theta & \cos \theta & 0\\0 & 0 & 0 & 1 \end{array} \right|; R_y = \left| \begin{array}{cccc} \cos \theta & 0 &-\sin \theta & 0 \\0 & 1 & 0 & 0\\ -\sin \theta & 0 & \cos \theta & 0\\0 & 0 & 0 & 1 \end{array} \right|; R_x = \left| \begin{array}{cccc} \\\cos \theta & -\sin \theta & 0 & 0\\ \sin \theta & \cos \theta & 0& 0\\ 0 & 0 & 1&0\\ 0 & 0 & 0 & 1 \end{array} \right|
\end{equation}

\setcounter{equation}{\value{MYtempeqncnt}}
\hrulefill
\vspace*{4pt}
\end{figure*}
The transformation matrix (M = translation + rotation) for each voxel is derived automatically by analyzing the difference between source and target data of the registration, which can again be described as a sequence of MAC operations. Note that the MRI and the PET images with reasonable accuracy have number of voxels more than $512\times 512 \times 50$. Therefore, a standard registration operation requires $> 13,10,7200$ MAC operations which can consume a significant amount of power and computation time. Similarly, other image analysis operations, such as object segmentation, feature extraction, can be decomposed into a set of sequential MAC operations. Hence, reduction in  power consumption during MAC operations is a significant challenge that has to be addressed in order to develop on-chip image analysis blocks.

Spurred on by Moore's law and CMOS technology scaling, the  general approach towards CAD has been to perform all mathematical  operations digitally using computers. While digital computation can provide high accuracy, the power consumption has been prohibitively high to prevent portable solutions. With CMOS technology scaling slowing down, there is an interest in analog signal processing (ASP) to perform mathematical computations in a low power fashion \cite{3}. A key enabler for ASP methodology is that most image processing algorithms require only 6-8 bits precision. ASP excels at approximate computing while consuming lower power than digital computing. Recent approaches to analog computation has been to use switches and capacitors as signal processing elements \cite{4,5,6,7} in advanced CMOS technologies. The matrix multiplier reported in \cite{7} performs calculations with 6b linearity while having an energy consumption of only 13fJ/operation at 1GHz.
 The work in \cite{4} presents switched capacitor multipliers and dividers. However, it uses amplifiers which is not power efficient, specially in advanced CMOS technologies. The works in \cite{5,6,7} present power efficient multipliers by  doing away with amplifiers and relying on only switched capacitors for signal processing.  These multipliers work in the charge domain and  are good solutions for high speed operations usually in the range of GHz of bandwidths. However, charge leakage presents a significant challenge to using switched-capacitor multipliers for medical signal processing in which computations can require only a few MHz of bandwidths. Thus, switched capacitor multipliers are not very suitable for medical image processing.  
 
In this letter, we present an alternate solution by performing multiplication and addition operations in time domain, which satisfies medical image processing requirements. The proposed technique has significantly higher immunity to charge leakage than switched capacitors  and can trade-off speed for power without sacrificing computational accuracy.  Time domain circuits are uniquely suitable for scaled CMOS technology. They are highly digital and hence can operate at low supply voltages in advanced CMOS technologies.  In addition, the quantization noise in time domain circuits is essentially transistor delay which reduces with technology scaling. Thus, the proposed architecture can be used for  a power efficient hardware implementation of the rigid registration block (see Fig. \ref{fig0}) which is a core element in PET/MRI modality fusion process. In addition, the proposed time domain architecture holds the promise of providing low power computational ability for a wide range of portable CAD systems.
 
 The rest of this letter is organized as follows: Section \ref{sec2} presents a brief review of existing analog-to-digital matrix multipliers, Section \ref{sec3} discusses the key idea behind the proposed time-domain matrix multiply-and-add operator, Section \ref{sec4} presents a CMOS circuit implementation and simulation results, while the conclusion and future research direction are brought up in Section \ref{sec5}.

\section{Review of Analog-to-Digital Matrix Multiplication and Addition}
\label{sec2}
A MAC operation is defined as 
\begin{eqnarray}
\addtocounter{equation}{1}
Y = \sum _{j=1}^N W_j X_j \label{eqn1}
\end{eqnarray}
where $X_j$ is  the input and  $W_j$ is the weight. 

Several analog signal processing techniques have been reported which accept an input $X_j$, perform the multiplication in (\ref{eqn1}) in analog domain and return a digital output $Y_j = W_jX_j$. A straightforward method is to use an operational amplifier to perform voltage domain multiplication and addition as described by (\ref{eqn1}). The operational amplifier needs to have a high gain to perform the MAC operation accurately. However, high gain operational amplifiers are power hungry and instead four-quadrant CMOS multipliers \cite{quadrant} are often used for approximate multiplication.  To reach higher energy efficiency, \cite{8} uses charge coupling to implement approximate analog matrix multipliers. \cite{9} uses low performance, thin-film transistors (TFT) to implement an approximate analog multiplier for processing data from sensors.

 More recently, there have been efforts to integrate analog-digital matrix multiplication (AD-MM) inside analog-to-digital converters (ADC) to lower power consumption. Capacitors and switches are used for passive multiplication in \cite{6,7,sar}. Charge domain passive multipliers achieve very low power consumption in advanced CMOS technology nodes as they do not use any active amplifiers which are power hungry. Addition can be done in current or charge domain in a low power fashion.

\section{Time Domain Matrix Operations}
\label{sec3}

While passive switched capacitor techniques are a good solution for low energy, approximate multipliers, they suffer from non-idealities due to charge leakage specially for advanced CMOS technologies which suffer from increased leakage. This problem is exacerbated at low speeds of operation, which implies that further lowering of energy consumption by reducing speed is  challenging for these techniques. In addition, often switched capacitor multipliers are used in conjunction with voltage domain ADCs which suffer from reduced dynamic range as the supply voltage is scaled down in advanced CMOS technologies.

To counter these limitations of switched capacitor multipliers, we propose a time domain multiplier. By shifting to time domain, sensitivity to charge leakage is greatly diminished and very low energy multipliers can be designed which are suitable for low bandwidth medical image processing. An added advantage is that quantization noise of time-domain multipliers come from transistor delay which reduces with technology scaling. Thus, in advanced CMOS technologies,  time-domain multipliers are better suited for low-bandwidth operations than switched capacitors.

A voltage-controlled oscillator (VCO) is a major building block of time domain multiplier. 
The quantized phase of a VCO can be written as
\begin{eqnarray}
\displaystyle \phi [k] = 2\pi K_{v} \int_{0}^{T_{int}} V_{in}(t) dt 
\label{eqn2}
\end{eqnarray}
where $V_{in}$ is the input to the VCO, $K_v$ is the VCO gain and $T_{int}$ is time over which the VCO phase is integrated. If $V_{in}$ changes slowly, $\phi [k]$ can be written as

\begin{eqnarray}
\displaystyle \phi [k] &=& 2\pi K_v \sum_{j=1}^N (t_{j}-t_{j-1}) V_{in}[j] \nonumber\\
&=& 2\pi K_v \sum_{j=1}^N \left(\Delta t_j\right) V_{in}[j]
\label{eqn3}
\end{eqnarray}
where $T_{int}=\displaystyle \sum_{j=1}^N \Delta t_j$ and $V_{in}[j] = V_{in}(t=t_j)$

(\ref{eqn3}) is analogous to (\ref{eqn1}) with $W_j \equiv 2\pi K_v (\Delta t_{j})$ and $X_j \equiv V_{in}[j]$.
Thus, a VCO can be used to perform a MAC operation in phase domain. The equivalent digital output of the MAC operation can be readily obtained by simply sampling the output of the VCO, without requiring a separate ADC as in the charge-domain matrix multipliers. The accumulation operation  is done in phase domain and is highly linear. Unlike charge-domain architectures, accumulation in phase domain comes without any additional power consumption.  Nonlinearity in phase domain MAC operation is primarily due to  nonlinearity in voltage-to-phase conversion. Increasing the integration time, $T_{int}$, allows reduction of VCO gain, $K_v$, which in turn increases VCO linearity. This is particularly suitable for medical image processing which does not require a high bandwidth, and hence, linearity of the VCO can be increased by lowering $K_v$ and increasing $T_{int}$.  As long as the VCO is oscillating, there is no leakage error in the phase value held by the VCO which is of importance  for low bandwidth medical signal processing.

\section{Matrix Multiplier Architecture}
\label{sec4}

\begin{figure}[tbh!]
	\centering
	{\includegraphics[width=0.44\textwidth]{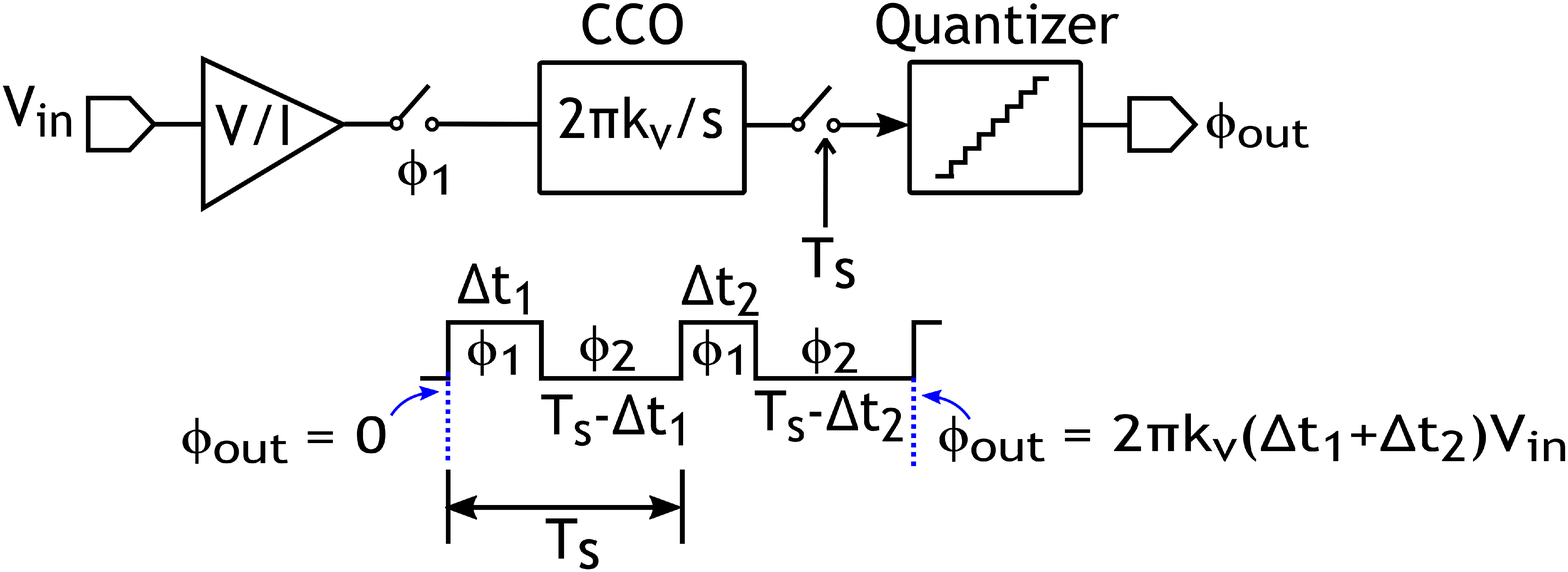}}
	\caption{Architecture and timing diagram of proposed time domain matrix multiplier. }
	\label{fig1a}
	\vspace{0mm}
\end{figure}

Fig. \ref{fig1a} shows the conceptual block diagram of the proposed time-domain matrix multiplier architecture along with its timing diagram. A voltage-to-current (V/I) converter drives a current-controlled oscillator (CCO) and the quantized phase output of the CCO holds the result of multiplication of $V_{in}$ and $2\pi K_v \phi_1$. The duration of $\phi_1$ in the $j-$th sampling period is  $\Delta t_j$ which is digitally controllable. Addition comes without any hardware cost as the CCO holds on to its phase which keeps accumulating with time. Fig. \ref{fig1a} illustrates how the proposed architecture performs MAC operation.

\begin{figure}[tbh!]
	\centering
	{\includegraphics[width=0.48\textwidth]{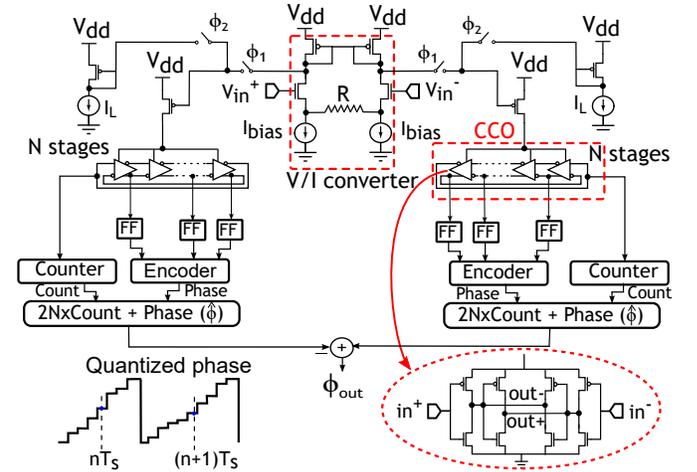}}
	\caption{Circuit schematic of  proposed time domain matrix multiplier. }
	\label{fig1b}
	\vspace{0mm}
\end{figure}

Fig. \ref{fig1b} shows  circuit implementation of the proposed architecture. The architecture is implemented in a differential fashion to suppress common-mode noise on the inputs as well as noise from supply and ground. The input signal is applied differentially to a V/I converter. The V/I converter drives two pseudo-differential CCOs during the phase $\phi_1$ and the two CCOs are run with a low current supply $I_L$ during the phase $\phi_2$.  The two CCOs are not stopped during $\phi_2$ to ensure that the accumulated phase held by them are not corrupted by leakage. By running the two CCOs at the same frequency during $\phi_2$, no phase is accumulated differentially at $\phi_2$.  The output of each CCO stage is latched by the sampling clock into a flip-flop (FF). At any given time, only one of the CCO stages is in a state of either a positive or negative transition. Thus, for an $N$-stage CCO, the instantaneous phase can be quantized into $2N$ levels between $(0,2\pi )$ corresponding to $N$ positive transitions and $N$ negative transitions. The matrix multiplier is highly digital and makes use of simple digital circuits to perform time domain analog signal processing.

As shown in Fig. \ref{fig1b}, CCO phase increases monotonically with time and wraps over after it crosses $2\pi$. A counter is used to store the number of times the VCO phase overflows over the period of integration. The total  phase at any time is given by $(2N\cdot Count + \hat{\phi} )$ where $\hat{\phi}$ is the instantaneous quantized phase.

\begin{figure}[tbh!]
\centering{\includegraphics[width=0.4\textwidth]{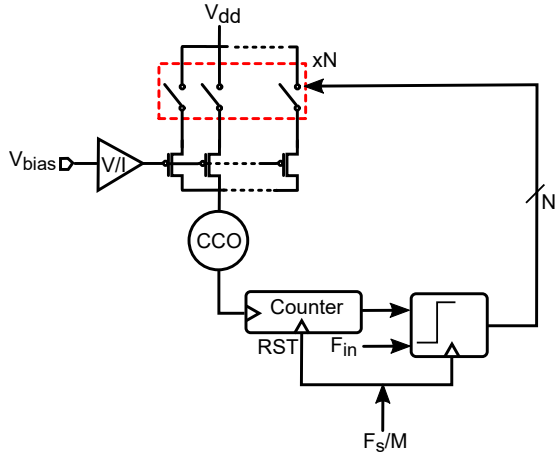}}
\caption{VCO gain tracking architecture. }
\label{fig2}
\vspace{0mm}
\end{figure}

Since VCO gain varies with process, voltage and temperature (PVT), the result of multiply-and-add will vary with PVT. Hence, background tracking is needed to correct for VCO gain variation. The V/I transconductance is given by $1/R$ for $g_mR \gg 1$, where $g_m$ is the transconductance of the V/I input transistors and $R$ is their source degeneration resistance. Thus, the V/I converter is relatively insensitive to PVT variations, and hence, resistance trimming is not required for multipliers with 6-8 bits accuracy. 
To track the CCO gain, width of the tail current source can be changed depending on the output of a counter which is clocked by the CCO as shown in Fig. \ref{fig2}. The counter output is monitored by a comparator which is clocked by a divided-down version of the sampling clock.  The counter is reset after every comparison. If the CCO is running too fast, the comparator will reduce the tail current source width  and vice versa. This ensures that the counter output is held equal to a preset value $F_{in}$ which sets the CCO free-running frequency. 

The background tracking technique can be applied to a reference multiplier and the comparator digital output word can be applied to the tail current source of all the CCOs. Process related mismatch between the tail current sources of the different CCOs will limit the accuracy to which the PVT sensitivity can be corrected. Fortunately, the matching accuracy is not very stringent as only 6-8 bits accuracy is required. In addition, CCO tail current devices are usually made large to reduce flicker noise. Large device size also reduces mismatches. For a large design with many multiplier cells, a few local copies of the reference multiplier can be distributed across the chip to account for gradient mismatches.

\begin{figure}[tbh!]
	\centering{\includegraphics[width=0.42\textwidth]{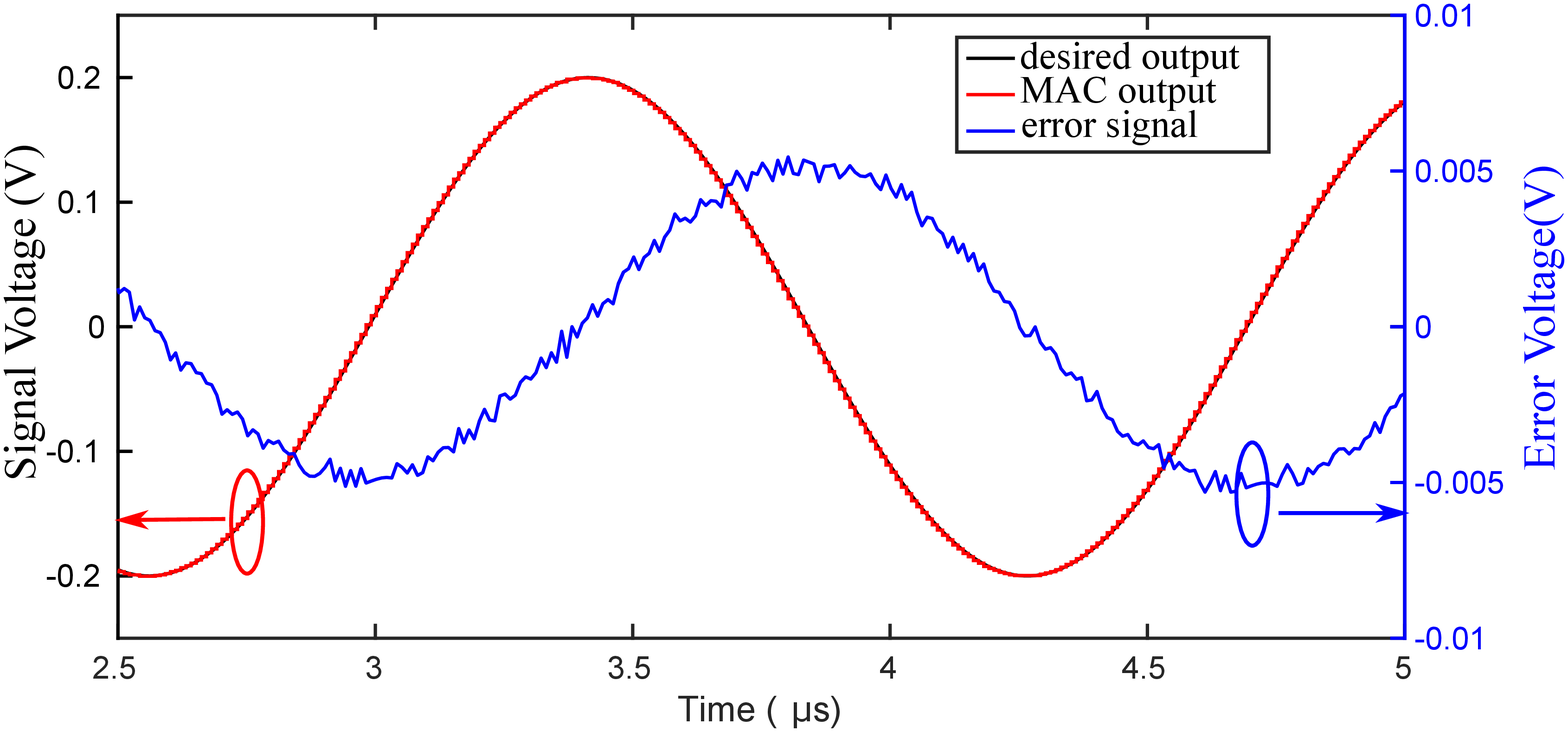}}
	\caption{Matrix multiply-and-add transient simulation. }
	\label{fig3}
	\vspace{0mm}
\end{figure}

A MAC cell was designed in 40nm CMOS process.  With a $400$mV differential input running at 0.6MHz, the MAC cell operates with 7-bit linearity. To test the accuracy of the proposed MAC cell, a row vector $(1\times512)$ is multiplied with a column vector $(512\times1)$. The row vector $\begin{bmatrix} x_0 & x_1 & x_2 & \cdots & x_{511} \end{bmatrix}$ is set to $\begin{bmatrix} 0 &\sin(\omega T_s) &\sin(2\omega T_s) &\cdots& \sin(511\omega T_s)\end{bmatrix}$ and all the elements in the column vector are set to $T_s$ where $T_s$ = 10ns. Fig. \ref{fig3} shows the simulated result for multiplication of the row vector and the column vector as well as the error between the output of the MAC operation and the desired output. It can be seen from Fig. \ref{fig3} that the proposed MAC cell has a low quantization error and has 7 bit linearity.

The MAC cell consumes $86\mu$W from a 1.1V supply while operating at 100 MHz. The energy efficiency of the proposed time-domain MAC cell is 2fJ/operation, compared to 968 fJ/operation energy efficiency of highly optimized digital matrix multipliers \cite{12}. Thus, the proposed MAC cell has more than 400X better energy efficiency than digital matrix multipliers. 

\section{Conclusion}
\label{sec5}

A time-domain architecture for performing multiplication-and-addition operations is presented in this letter.  The proposed architecture exploits the low bandwidth and not-too stringent accuracy requirement of medical image processing algorithms  to achieve drastic increase in energy efficiency compared to digital matrix multipliers. The architecture is highly digital and suitable for advanced CMOS technologies. The proposed architecture can act as an enabler for developing portable hardware solutions for 2D/3D image analysis.

\end{document}